\documentclass[nonacm,sigconf]{acmart}

\AtBeginDocument{%
  \providecommand\BibTeX{{%
    \normalfont B\kern-0.5em{\scshape i\kern-0.25em b}\kern-0.8em\TeX}}}

\setcopyright{acmcopyright}
\copyrightyear{2024}
\acmYear{2024}
\acmDOI{XXXXXXX.XXXXXXX}

\acmConference[]{}{}{}
\setcopyright{none}
\usepackage[utf8]{inputenc}
\usepackage{algorithmic}
\usepackage{graphicx}
\usepackage{textcomp}
\usepackage{xcolor}

\usepackage{textcomp}
\usepackage{xcolor}
\usepackage{hyperref}
\usepackage{cleveref}
\usepackage{xspace}
\usepackage{paralist}
\usepackage{multirow}
\usepackage{comment}

\usepackage[tikz]{bclogo}
\usepackage{pgfplots}
\pgfplotsset{compat=1.17}
\usepackage{subcaption}
\usepackage[export]{adjustbox}

\newcommand{\company}{Microsoft}%{Microsoft\xspace}
\newcommand{\etal}{\emph{et al.}\xspace}

\renewcommand\footnotetextcopyrightpermission[1]{} % removes footnote with conference information in first column
\begin{document}

\title{Dependency Aware Incident Linking in Large Cloud Systems}

\author{Supriyo Ghosh, Karish Grover, Jimmy Wong, Chetan Bansal, Rakesh Namineni, Mohit Verma, Saravan Rajmohan}
\email{{supriyoghosh, t-kgrover, jimmywong, chetanb, rakesh.namineni, mverma, saravan.rajmohan}@microsoft.com}
\affiliation{%
   \institution{Microsoft}
  \streetaddress{}
   \city{}
   \country{}
}

\newcommand\chetan[1]{\textcolor{blue}{Chetan: #1}}

\renewcommand{\shortauthors}{Ghosh et al.}

\begin{abstract}
Despite significant reliability efforts, large-scale cloud services inevitably experience production incidents that can significantly impact service availability and customer's satisfaction. Worse, in many cases one incident can lead to multiple downstream failures due to cascading effects that creates several related incidents across different dependent services. Often time On-call Engineers (OCEs) examine these incidents in silos that lead to significant amount of manual toil and increase the overall time-to-mitigate incidents. Therefore, developing efficient incident linking models is of paramount importance for grouping related incidents into clusters so as to quickly resolve major outages and reduce on-call fatigue. Existing incident linking methods mostly leverages textual and contextual  information of incidents (e.g., title, description, severity, impacted components), thus failing to leverage the inter-dependencies between services. 
In this paper, we propose the \textbf{d}ependency-aware \textbf{i}ncident \textbf{link}ing (DiLink) framework which leverages both textual and service dependency graph information to improve the accuracy and coverage of incident links not only coming from same service, but also from different services and workloads. Furthermore, we propose a novel method to align the embeddings of multi-modal (i.e., textual and graphical) data using Orthogonal Procrustes. Extensive experimental results on real-world incidents from 5 workloads of \company{} demonstrate that our alignment method has an F1-score of 0.96 (14\% gain over current state-of-the-art methods). We are also in the process of deploying this solution across 610 services from these 5 workloads for continuously supporting OCEs improving incident management and reducing manual toil.
\end{abstract}


\maketitle

\section{Introduction} \label{intro}
Large-scale cloud operators (e.g., Google, Microsoft, Amazon) run tens of thousands of services with highly complex architecture and inter-dependencies between the services. Despite significant reliability efforts to ensure continuous availability of services, production incidents (e.g., unplanned interruptions or performance degradation) are inevitable in large-scale cloud systems, which adversely impacts the customer satisfaction. These incidents can be extremely expensive in terms of customer impact, revenue loss via violation of service-level agreements, and manual toil required from On-call engineers (OCEs) to resolve them. For example, the estimated cost for one hour of service downtime for Amazon on a major shopping day is approximately US\$100 million \cite{amazon-100-million}. 

In many cases, due to inter-dependencies among services one failure can cause cascading effects that propagate error to down-streaming services. As engineers from every service team set up their own automated watchdogs with specific alert rules for faster detection of incidents, such cascading effect leads to many alerts being reported from different services within a short span of time, which is referred as alert storm. These problems are highly impactful but notoriously challenging to handle without proper domain expertise and knowledge of inter-dependencies among services. OCEs often inspect these related incidents in silos leading to higher engagement of engineering resources, repetitive efforts and manual toil, and delay in recovering service health. On the other hand, multiple independent faults may also arise within a short time period, which should be examined by OCEs separately in a timely fashion. Therefore, accurately clustering similar and related incidents is of paramount importance to reduce the burden of OCEs and ensuring reliability of cloud systems. In addition, an accurate incident linking model can also provide the OCEs with relevant clues and evidence to check the root causes of the incidents quickly.

Given the practical importance, several recent works studied the incident aggregation \cite{chen2021graph,lin2014unveiling,zhao2020understanding} and incident linking \cite{chen2020identifying,chen2022online,gu2020efficient} problems. While the goal for incident aggregation is to cluster incidents caused by the same failure, incident linking determines whether two incidents are similar (related, duplicate or responsible). Gu \etal \cite{gu2020efficient} proposed a transfer learning strategy to effectively link customer and monitor reported incidents. Existing incident linking models mainly focused on textual and contextual information (e.g., title, description, severity, impacted components) of incidents that generally performs poorly when the incidents are coming from different services and workloads. To address this challenge, it is important to encapsulate the dependency relationship information among different services within the incident linking model. Chen \etal \cite{chen2020identifying} recently proposed a data-driven approach called LiDAR that computes the similarity from textual information and component graph separately and use a convex combination of them to generate the similarity score. However, as shown in our experimental results, learning the similarity between textual and graphical data separately can only provide marginal benifit.

Therefore, we propose \textbf{d}ependency-aware \textbf{i}ncident \textbf{link}ing (DiLink) framework by leveraging dependency graph information along with textual and contextual information of incidents for improving the accuracy of incident linking process, especially for cross-service and cross-workload incidents. Our goal is to answer the following research questions:
(1) Whether augmenting textual information with dependency graph data can improve accuracy of incident linking model? 
(2) How to efficiently combine multi-modal (i.e., textual and graphical) data for near-optimal incident linking accuracy?
(3) How the linking accuracy varies between within-service, cross-service and cross-workload incident pairs? and
(4) How completeness of the dependency graph affects the incident linking accuracy?

To answer these questions, we first develop a dependency graph among 610 services coming from 5 workloads in \company{} using service meta-data information and historical incident links. The dependency graph has 610 nodes (representing the services) and more than 5500 edges. For each service, we create a sub-graph by considering its 3 hop neighbours (both from incoming and outgoing edges). While generating the embeddings of an incident in DiLink, we compute both the embeddings from textual components (e.g., title, topology, monitor id, failure type and owning service name) and graphical components (i.e., sub-graph corresponding to the owning service of the incident). However, due to misalignment of embeddings, we experimentally observe that a simple concatenation of textual and graphical embeddings leads to poor performance. To address this challenge, we propose a novel alignment method using the concept of Orthogonal Procrustes \cite{gower2004procrustes} from linear algebra. To train the model, we generate more than 1 million triplets with an anchor incident and its corresponding related (i.e., positive example) and non-related (i.e., negative example) incidents using 9 months of historical data from 2022. Finally, we evaluate the performance on last 3 months of incidents from 2022 and demonstrate that DiLink provides a significant 14\% gain in accuracy and F1-Score in comparison to state-of-the-art methods. In summary, our key contributions are as follows:  
\begin{itemize} 
    \item We thoroughly investigate whether augmenting textual information with dependency graph data can improve the accuracy of incident link predictions, especially when incidents are reported from different services and workloads. 
    \item We propose DiLink, that employs Orthogonal Procrustes method from linear algebra for alignment of textual and graphical embeddings. 
    \item With extensive experimental results on incidents from 610 services, we demonstrate that DiLink can achieve an F1-score of more than 0.96, a significant 14\% gain over state-of-the-art methods. In addition, we are in process of deploying our model in real production scenario for 5 workloads within \company{}.
\end{itemize}

The remainder of this paper is as follows: an overview on the incident management and linking procedure is provided in Section \ref{background}. In Section \ref{methods}, we illustrate our proposed framework for incident linking in large-scale cloud systems. In Section \ref{experiments}, we demonstrate the efficacy of our proposed method experimentally on real-world incident dataset. In Section \ref{deployment}, we provide the details of deployment of our model. We then summarize the lessons learnt in Section \ref{lessons}, related work in Section \ref{related}, and conclude our work in Section \ref{conclusion}.

\section{Background} \label{background}
In this section, we provide an overview of incident management and incident linking process within \company{}.

\subsection{Incident management}
\begin{figure}[ht]
    \centering
    \includegraphics[width=0.45\textwidth]{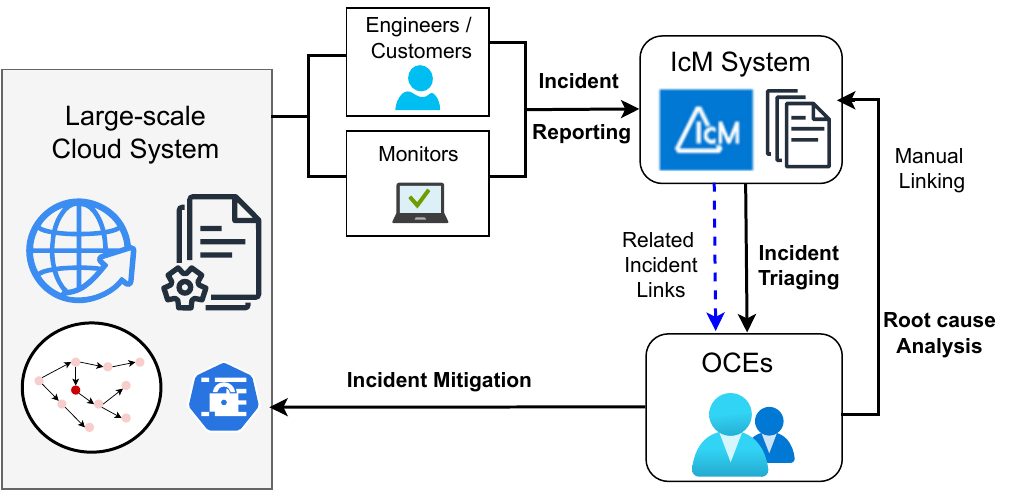}
    % \vspace{-0.1in}
    \caption{Incident management lifecycle.}
    \Description{Incident management lifecycle.}
    \vspace{-0.1in}
    \label{fig:lifecycle}
\end{figure}

Every production incident information within \company{} is recorded in a centralized incident management (IcM) portal. Lifecycle of an investigated incident typically has four stages as shown in Figure~\ref{fig:lifecycle}. 
\begin{enumerate}
    \item \textbf{Detection:} To quickly detect the failures within a service, engineers usually design automated watchdogs that continuously monitors the system health and report an incident in IcM when an anomaly is detected, which is referred as monitor-reported incidents (MRIs). Incidents can also be reported either by internal or external customers of a given service, which is referred as customer-reported incidents (CRIs). In this study, we primarily looked at relationships between MRIs.  
    \item \textbf{Triaging:} Once an incident is reported, a team of OCEs quickly investigation the details and route the ticket to appropriate OCEs. This incident triaging process can take multiple rounds in some cases to reach the accurate responsible team. 
    \item \textbf{Diagnosis:} Once an incident is routed to the appropriate team of OCEs, multiple iterations of back and forth communication takes place between the OCEs inspecting different aspects to understand key root causes of the failure. When required, the OCEs manually identify similar incidents based on diagnosis information and log the linking information in IcM. 
    \item \textbf{Mitigation:} Once the root cause is identified, several actions are taken to mitigate the problem and to recover the service health.
\end{enumerate}

\subsection{Incidents considered in our study}
We have considered historical incident links from 5 following workloads of \company{}:
\begin{itemize}
\item Workload-1 (Exchange): It is a large-scale back-end service for mail and calendar and used by more than 400 million users. It includes communication and organizational features such as email hosting, shared calendars, and other task management components. %(Exchange)
\item Workload-2 (EOP): This is a cloud-based filtering service that protects services of Workload-1 against spam and malware. %(EOP): Exchange Online Protection.
\item Workload-3 (Outlook): It is the frontend of Workload-1 that can be used as stand-alone application including functionalities such as calendaring, task managing and contact managing. % (Outlook)
\item Workload-4 (Skype-Teams): It is a complex web-scale distributed service powering messaging, calling and meeting services for the real time communication productivity application from \company{} and used by more than 250 million users worldwide. % (Skype-Teams)
\item Workload-5 (Intelligent Conversation and Communication Cloud [IC3]): It supports the backend of Workload-4 and provides functionalities for real-time intelligent communication. % (IC3)
\end{itemize}
We considered only MRI incidents (97.7\% of all reported incidents) from  January 01, 2022 to January 01, 2023 in our study. The incident severity level within \company{} ranges from 1 to 4, but we filter out severity 4 incidents as those are usually noisy and transient in nature. 

\begin{figure}[!htb]
    \centering
    \scalebox{0.9}{\input{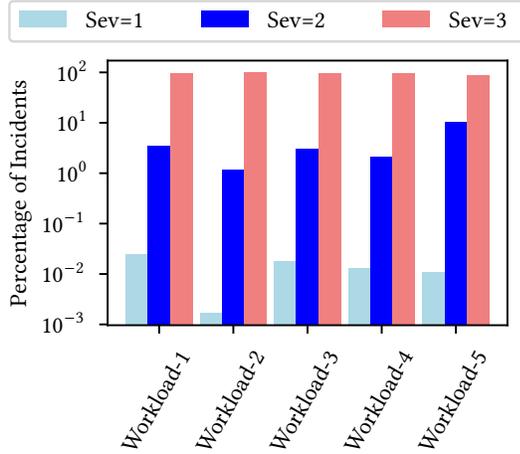}}
    \vspace{-0.1in}
    \caption{Distribution of different severity incidents.}
    \Description{Distribution of different severity incidents.}
    \vspace{-0.1in}
    \label{fig:incounts}
\end{figure}

\subsection{Incident Linking}

In the IcM system within \company{}, OCEs manually linked incidents with variety of types depending on the nature of the links. We provide a few examples in Table~\ref{tbl:incident_links}. The incident links are typically duplicate or related or responsible, whereas the parent and child incidents can either come from same team, or different teams within a particular workload, or even reported from different workloads.

\paragraph{\textbf{Duplicate links.}} If two linked incidents originates from same failure or malfunction of certain component, we refer to them as duplicate links.  As the incidents can be reported by different sources (e.g., customers, engineers, monitors), often time multiple incidents are reported for the same failure. In some cases, even multiple monitors tracking the same health metric, can report multiple duplicate incidents for a particular anomaly. For example, the first pair of incidents in Table~\ref{tbl:incident_links} are reported for the failure of a particular endpoint. In our case, first incident refer to the parent incident which is reported and investigated by the OCEs earlier. Note that for both the incidents, a particular workload in a particular location is impacted due to failure of endpoints which creates both anomaly and connectivity degradation, thus these two are duplicate incidents.

\paragraph{\textbf{Related links.}} 
If the nature and descriptions of two incidents are different but both of them are triggered from a common fault, we refer to them as related incident links. These incidents are neither duplicate nor one causes other incident sequentially. The second pair of incidents in Table~\ref{tbl:incident_links} is an example of related linked incidents where the Workload-2 in a particular region is impacted due to some failure which causes CPU or memory overload issue as well as network routing is impacted. 

\paragraph{\textbf{Responsible links.}} When one failure has a cascading effect that impacts other services due to complex dependencies among each other, one incident sequentially lead to other incidents, which are referred as responsible incident links.
For example, if the database in a particular region is impacted due to storage or networking issue, it leads to anomalies in products leveraging that particular database. These incidents are notoriously challenging to link together unless OCEs have a deeper understanding and expertise of system functionalities and dependencies. The third incident pair in Table~\ref{tbl:incident_links} is a challenging example, where due to some bugs, the calendar event information in Workload-4 is impacted that leads to another issue in Workload-1 as the availability of active directory information of the users dropped considerably. From the description of incidents, although these two issues seem independent, in reality these two incidents happened sequentially due to cascading effect.

\paragraph{\textbf{Cross team incident links.}} 
In our current IcM system, a large number of linked incidents are generated from different teams. There could be two types of cross team incidents: (1) when two incidents arise from different teams but from the same workload, we refer to them as \emph{cross-team} incidents; and (2) when two incidents arise from different workloads, we refer to them as \emph{cross-workload} incident links. These links are usually challenging to detect and the textual description might include different aspect of the teams and workloads that increases the distance in embedding space and reduces similarity score. We are particularly interested in these incident links where dependency information are primarily important to identify true links between these incidents. In Table~\ref{tbl:incident_links}, the second pair of incidents is an example of cross-team link where two different teams in Workload-2 are impacted. On the other hand, the third pair of incidents is an example of cross-workload incident link as two different teams in two different workloads are impacted. 
\begin{table*}[htb]
\caption{Example of linked incidents}
\label{tbl:incident_links}
\begin{center}
\begin{tabular}{|p{2.0cm} | p{1.5cm} | p{9.0cm} | c|} 
\hline
{Type} & ID & Title & LinkType \\
\hline
\multirow{2}{*}{Within-Team} & Incident-1 &  Anomaly Detected for Endpoint(s) of Workload-3 at locationX & \multirow{2}{*}{Duplicate}\\ 
\cline{2-3}
 & Incident-2 & Connectivity Degradation for Endpoint(s) in Workload-3 at locationX & \\ 
\hline
\hline
\multirow{2}{*}{Cross-Team} & Incident-1 & CPU or Memory overload in CountryX and RegionY & \multirow{2}{*}{Related}\\ 
\cline{2-3}
 & Incident-2 & Issues in Workload-2 in CountryX causing routing problems for RegionY & \\ 
\hline
\hline
\multirow{2}{*}{Cross-Workload} & Incident-1 & User calendar events view in Workload-4 is unhealthy & \multirow{2}{*}{Responsible}\\ 
\cline{2-3}
 & Incident-2 & The directory availability in a target forest in Workload-1 has dropped below threshold value & \\ 
\hline
\end{tabular}
\end{center}
\end{table*}

\begin{figure}[!htb]
    \centering
    \scalebox{0.9}{\input{Figures/Incident_time.pgf}}
    \vspace{-0.1in}
    \caption{Time to mitigate (TTM) for linked vs. all incidents \textmd{(Y-axis shows the normalized TTM value with the median of TTM of all incidents as 1.)}.}
    \Description{Time to mitigate (TTM) for linked vs. all incidents.}
    \vspace{-0.1in}
    \label{fig:intime}
\end{figure}

In Figure~\ref{fig:intime}, we compare the median time-to-mitigate (TTM) for linked MRI incidents across 5 workloads against the median TTM for all incidents. The Y-axis shows the normalized TTM value with the median of TTM of all incidents across 5 workloads considered as 1. The TTM for linked incidents are typically higher because it takes time for OCEs to manually link them and often time different OCEs investigate related incidents in silos.
Therefore, automated incident linking is a practically challenging and important
task in quickly resolving incidents and reducing manual toil for OCEs. To correctly predict the links among incidents, we not only need the semantics of textual description of incidents, but also need to model the system dynamics and dependencies among services. Therefore, we propose to augment the textual description of incidents with service dependency graph information to accurately predict the incident links.

\section{Methodology} \label{methods}
We propose DiLink to accurately predict related incidents reported from different services and workloads. In this
section, we begin by introducing the overall architecture of the proposed model, that is composed of two modules representing the textual
descriptions and the dependency graph information, respectively.
Then we explain each of the modules in depth and describe the method for alignment of multi-modal embeddings, followed by details of model training and real-time inferencing procedure.

\subsection{Overview of architecture}
\begin{figure*}[!htb]
    % \centering
    \includegraphics[width=0.85\textwidth, right]{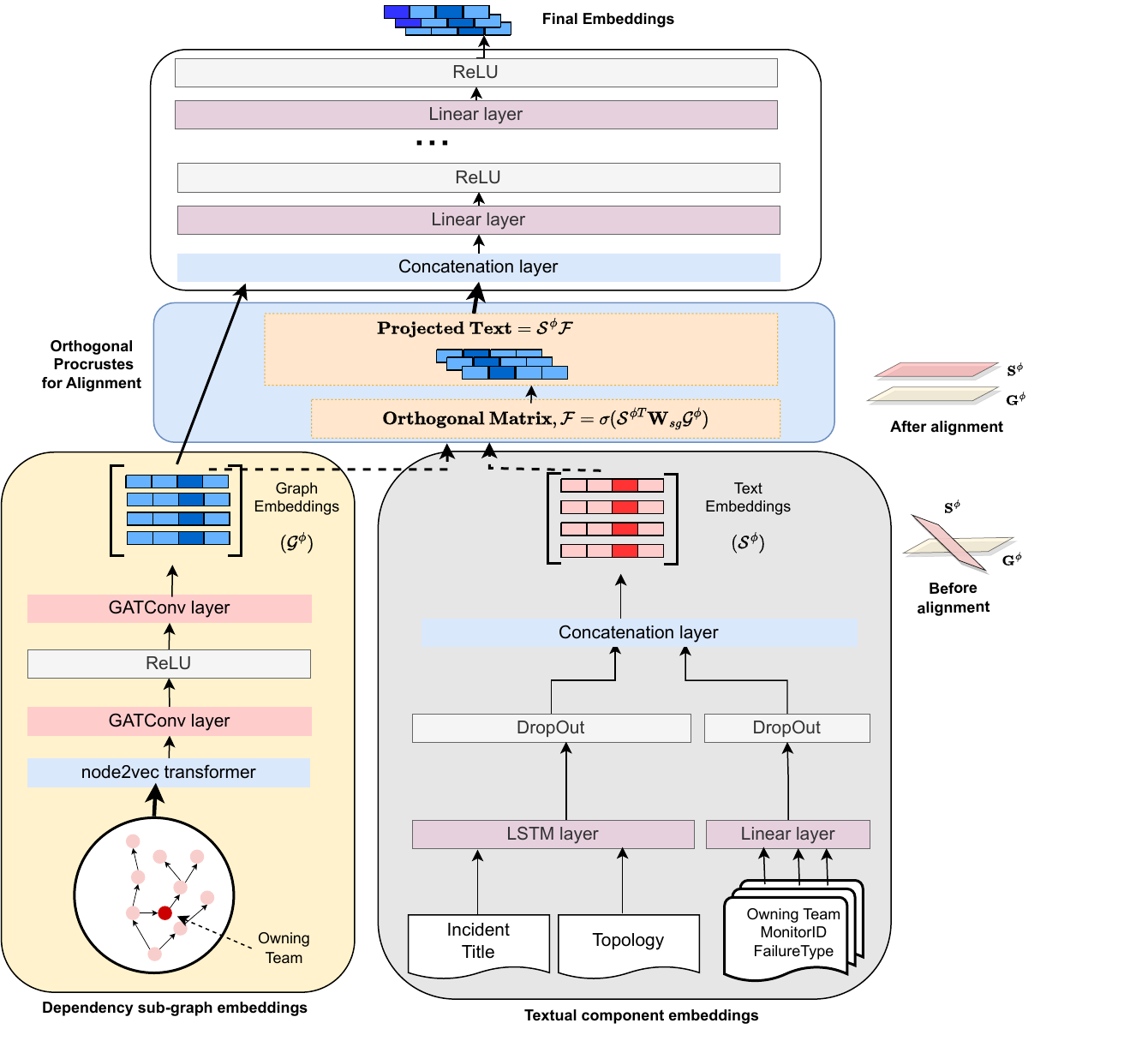}
    \vspace{-0.2in}
    \caption{Overall architecture of DiLink.}
    \Description{Overall architecture of DiLink.}
    \vspace{-0.1in}
    \label{fig:architecture}
\end{figure*}

The overall structure of our proposed DiLink method is shown in Figure~\ref{fig:architecture}. 
We consider both the textual description (i.e., title, topology), categorical information (i.e., monitorID, failure type and owning teamID) and service structural information (i.e., dependency graph representation) from each incident to learn the relationship between incidents. These features are fed into two module: the textual embedding module to extract information from textual and categorical information and graph embedding module to extract information from dependency graph. The textual module is trained to map the textual information into high dimensional semantic embedding space. The graph module is learned to be embedded into low dimensional latent space to represent the correlation between different services. We then employ Orthogonal Procrustes method to project the textual embeddings into graph embedding space so that both the aligned embeddings can be concatenated. Afterwards, the concatenated embeddings are passed through a set of Linear layer and ReLU \cite{agarap2018deep} activation function to get the final embeddings that are used to compute the similarity score between incidents. 

\subsection{Modeling textual information} % Karish
An incident title describes the symptom of the incident, including the location, impacted customers, service health information, etc. In addition, topological information includes details about impacted region, machine information (e.g., machine name, data center name, device group name), deployment ring information, and impacted service information. These information are essential to understand the potential links with other incidents. We learn the relationship between two incidents by learning the semantic representation of these textual information. To learn mapping from texts in natural language into a representative numerical vector in the latent space, we first use a TF-IDF (Term Frequency times Inverse Document Frequency) vectorizer. These numerical values are then passed through a LSTM (long short-term memory) \cite{hochreiter1997long} layer followed by a dropout layer to get the final embeddings.
In addition, we used 3 categorical features (incident owning team name, unique identification number of the monitor reported the incident and failure type reported by the monitor) which carry crucial information for identifying incident relationship. To get the embeddings for these features, we used TF-IDF vectorizer followed by a neural network with fully-connected layer. Finally, the embeddings from the textual description and categorical information are concatenated together to obtain a $D$-dimensional embedding from textual module. Finally, we pass this $D$-dimensional embedding through another linear layer that project the output into a $\hat{D}$-dimensional embedding which represents the learned vector for the textual description of each incident.

\subsection{Modeling Dependency graph information}

\subsubsection{Dependency graph construction}
Understanding the dependencies among different services in a hyperscale cloud system is a non-trivial task due to complex interaction between vast number of micro-services. Our goal is to generate a dependency graph $\mathcal{G}=<\mathcal{V},\mathcal{E}>$, where $\mathcal{V}$ denotes the set of nodes each represents a unique service, and $\mathcal{E}$ denotes the set of edges representing the dependency relationship between services. To generate the dependency graph, we first leverage the system metadata information available in dependency tracking system (DTS) tool within \company{}. This tool curates several information to link the services including shared subscription information, shared resource information, logs of service communication using domain name system (DNS). From this, we obtain a partial dependency graph by adding a link from source to dependent service for each record available in DTS tool. We then augment this partial dependency graph with additional edges generated from historical incident relationships. For each related link, we have the information regarding the owning service of both the parent and child incident. Therefore, for each related link, we create a new edge between owning service of parent incident to owning service of child incident in the dependency graph. By employing dependency links from these two data sources, we obtain our final dependency graph $\mathcal{G}$, which has 610 nodes and more than 5,500 edges. 

\subsubsection{Encoding dependency graph} To represent the semantics of dependency graph, we generate a sub-graph for each owning service from the global dependency graph by taking 3-hop distance neighbour nodes (for both incoming and outgoing edges). These sub-graphs are then converted into low-dimensional vector space using node2vec \cite{grover2016node2vec} graph transformer with random walks (walk length = 20 and number of walks = 100) for learning continuous feature representation of nodes in the sub-graph. Finally, these represented vectors are sent through graph representation networks for learning the low-dimensional embeddings of nodes from graph embedding module that uncovers the relationship between different sub-graph architectures. We used 3 well-known graph representation network to obtain the final sub-graph embeddings.

\paragraph{Graph convolution network (GCN) \cite{kipf2016semi}:} GCN is a popular scalable semi-supervised learning approach for classification of nodes in the graph, motivated from a first-order approximation of spectral graph convolutions \cite{hammond2011wavelets}.

\paragraph{Graph attention network (GAT) \cite{velivckovic2017graph}: } GAT employs masked self-attention layers, inspired by canonical transformer based attention network \cite{vaswani2017attention}, in which nodes are able to attend over their neighborhoods’ features, and therefore, able to assign different importances to nodes of a same neighborhood.

\paragraph{GraphSAGE \cite{hamilton2017inductive}: } GraphSAGE is a general inductive learning framework that can efficiently generate embeddings for previously unseen nodes by leveraging current nodes' features. Rather than learning embeddings for each node separately, it generates embeddings by sampling and aggregating features from a node’s local neighborhood, thus generalizes well to unseen nodes.

\subsection{Alignment of multi-modal embeddings}
As shown experimentally, a simple concatenation of embeddings from textual and graphical module leads to poor performance due to misalignment of vector representations. Therefore, we propose to use Orthogonal Procrustes \cite{gower2004procrustes} method from linear algebra for projecting the text embeddings to graph embedding space. Let $\mathcal{S}^{\phi}$ denote a $N\times \hat{D}$ dimensional vector representing the textual embeddings for a batch of $N$ incidents. Let $\mathcal{G}^{\phi}$ denote a $N\times \hat{D}$ dimensional graph embedding vector. The goal is to find the nearest orthogonal matrix, $R$ to a given matrix $M={\mathcal{S}^{\phi}}^T \mathcal{G}^{\phi}$, by solving the following approximation problem: $$\min_R ||R-M||_F, \hspace{0.5in} \text{Such that }\hspace{0.1in}  R^T R = I$$
We use singular value decomposition (SVD) method \cite{klema1980singular} for obtaining the $\hat{D} \times \hat{D}$ dimensional orthogonal matrix $R$, and then $\hat{\mathcal{S}}^{\phi} = \mathcal{S}^{\phi} R$ provides the $N \times \hat{D}$ dimensional projected text embeddings which closely aligns with graph embeddings, $\mathcal{G}^\phi$. Finally, we concatenate the graph embeddings, $\mathcal{G}^{\phi}$ and projected text embeddings, $\hat{\mathcal{S}}^{\phi}$ together and pass the joint embeddings through a set of linear neural networks and non-linear ReLU activation functions to get the final learned representation of an incident.

\subsection{Model training}
To learn the similarity score between two incidents, the final joint embeddings are trained based on a Siamese neural network \cite{koch2015siamese}, which is a popular state-of-the-art method for learning entity matching task. It has two twin neural networks, with identical structure and sharing the same set of parameters and weights \cite{kim2014convolutional}. Given a pair of incidents with their known relationship, two networks separately learns the vector representation of two incidents. This structure learns the embeddings in deep layer and places semantically similar features closer to each other. Finally, we train the Siamese network in a supervised learning fashion in accordance with given incident relationship labels. To train the model, we generates triplet of incidents with an anchor incident along with its corresponding positive and negative related incidents, and employ the following triplet loss function:
$$ Loss = \max (d(a_e, p_e) - d(a_e, n_e) + M, 0)  $$
where $d(a_e, p_e)$ and $d(a_e, n_e)$ denote the distance from anchor incident to positive and negative related incidents, respectively, and $M$ represent a non-negative margin value.

\subsection{Online prediction of links}
Once the Siamese network model is trained, we deploy the model (refer to Section \ref{deployment} for details of model deployment) and predict the relationship between a pair of incidents in an online fashion. When an incident is reported, we retrieve a set of incidents from $\Delta t$ lookback time period, and for each retrieved incident, we pass its embeddings along with reported incident embeddings to the trained Siamese model to get the similarity score. If the similarity score is less than a pre-defined tuned threshold value, then we classify it as a linked incidents.
\section{Empirical Analysis} \label{experiments}
We begin this section by summarizing the dataset, experimental set up and baseline methods, followed by demonstrating empirical results.  

\begin{table*}[t]
\centering
\caption{Effectiveness of different approaches}
\label{tbl:performance}
\begin{tabular}{p{2.0cm}p{2.5cm}p{2.0cm}p{2.0cm}p{2.0cm}c}
\hline
% Type, Method, Precision, Recall, Accuracy
\multicolumn{1}{c}{\multirow{1}{*}{Type}}  & \multicolumn{1}{c}{\multirow{1}{*}{Model}} & \multicolumn{1}{c}{Precision} & \multicolumn{1}{c}{Recall} & \multicolumn{1}{c}{F1-Score} & \multicolumn{1}{c}{Accuracy}   \\ \hline
\multirow{5}{*}{Overall} & Baseline & 0.765 $\pm$ 0.001 & 0.896 $\pm$ 0.0002 & 0.825 $\pm$ 0.0007 & 0.81 $\pm$ 0.0008 \\
 & Concatenation & 0.72 $\pm$ 0.002 & 0.908 $\pm$ 0.0002 & 0.803 $\pm$ 0.001 & 0.777 $\pm$ 0.002 \\
 & LiDAR & 0.737 $\pm$ 0.001 & 0.961 $\pm$ 0.0001 & 0.834 $\pm$ 0.001 & 0.81 $\pm$ 0.001 \\
 & DiLink-GCN & 0.93 $\pm$ 0.007 & 0.95 $\pm$ 0.008 & 0.94 $\pm$ 0.006 & 0.94 $\pm$ 0.007 \\
 & DiLink-GSAGE & 0.94 $\pm$ 0.023 & 0.956 $\pm$ 0.001 & 0.948 $\pm$ 0.01 & 0.947 $\pm$ 0.013 \\
 & DiLink-GAT & \textbf{0.968 $\pm$ 0.013} & \textbf{0.959 $\pm$ 0.002} & \textbf{0.963 $\pm$ 0.006} & \textbf{0.963 $\pm$ 0.006} \\
\hline       
\multirow{5}{*}{Within service} & Baseline & 0.59 $\pm$ 0.004 & 0.98 $\pm$ 0.0002 & 0.736 $\pm$ 0.003 & 0.615 $\pm$ 0.005 \\
 & Concatenation & 0.595 $\pm$ 0.006 & 0.94 $\pm$ 0.0004 & 0.728 $\pm$ 0.005 & 0.617 $\pm$ 0.005 \\
& LiDAR & 0.574 $\pm$ 0.008 & 0.973 $\pm$ 0.0005 & 0.722 $\pm$ 0.006 & 0.588 $\pm$ 0.007 \\
& DiLink-GCN & 0.96 $\pm$ 0.008 & 0.978 $\pm$ 0.01 & 0.97 $\pm$ 0.01 & 0.967 $\pm$ 0.01 \\
 & DiLink-GSAGE & 0.93 $\pm$ 0.034 & \textbf{0.998 $\pm$ 0.001} & 0.962 $\pm$ 0.019 & 0.956 $\pm$ 0.02 \\
 & DiLink-GAT & \textbf{0.978 $\pm$ 0.012} & 0.978 $\pm$ 0.004 & \textbf{0.978 $\pm$ 0.005} & \textbf{0.976 $\pm$ 0.005}\\
\hline
\multirow{5}{*}{Cross service} & Baseline & 0.793 $\pm$ 0.001 & 0.91 $\pm$ 0.0003 & 0.847 $\pm$ 0.0006 & 0.836 $\pm$ 0.0005 \\
 & Concatenation & 0.732 $\pm$ 0.002 & 0.927 $\pm$ 0.0002 & 0.818 $\pm$ 0.001 & 0.794 $\pm$ 0.002 \\
& LiDAR & 0.76 $\pm$ 0.001 & 0.976 $\pm$ 0.0001 & 0.854 $\pm$ 0.0006 & 0.834 $\pm$ 0.0006 \\
 & DiLink-GCN & 0.927 $\pm$ 0.007 & 0.944 $\pm$ 0.007 & 0.935 $\pm$ 0.006 & 0.935 $\pm$ 0.006 \\
 & DiLink-GSAGE &  0.94 $\pm$ 0.022 & 0.949 $\pm$ 0.001 & 0.945 $\pm$ 0.012 & \textbf{0.945 $\pm$ 0.012} \\
 & DiLink-GAT & \textbf{0.97 $\pm$ 0.013} & \textbf{0.956 $\pm$ 0.002} & \textbf{0.963 $\pm$ 0.006} & 0.927 $\pm$ 0.007 \\
\hline
\multirow{5}{*}{Cross Workload} & Baseline & 0.834 $\pm$ 0.013 & 0.672 $\pm$ 0.0005 & 0.745 $\pm$ 0.005 & 0.787 $\pm$ 0.003 \\
 & Concatenation & 0.83 $\pm$ 0.001 & 0.697 $\pm$ 0.001 & 0.757 $\pm$ 0.004 & 0.795 $\pm$ 0.003\\
& LiDAR & 0.796 $\pm$ 0.007 & 0.809 $\pm$ 0.0008 & 0.803 $\pm$ 0.003 & 0.819 $\pm$ 0.004\\
& DiLink-GCN & 0.93 $\pm$ 0.01 & 0.967 $\pm$ 0.01 & 0.947 $\pm$ 0.01 & 0.95 $\pm$ 0.01 \\
 & DiLink-GSAGE & \textbf{0.939 $\pm$ 0.02} & \textbf{0.97 $\pm$ 0.003} & \textbf{0.955 $\pm$ 0.012} & \textbf{0.958 $\pm$ 0.011} \\
 & DiLink-GAT & 0.937 $\pm$ 0.016 & 0.955 $\pm$ 0.001 & 0.946 $\pm$ 0.008 & 0.95 $\pm$ 0.008 \\
\hline
\end{tabular}
\vspace{-0.1in}
\end{table*}

\subsection{Experimental settings}
\subsubsection{Dataset}
Our proposed incident linking framework is evaluated on incidents coming from 610 services of \company{}. We collected information for all the incidents and links of 2022 from 5 workloads. We then split the data set into training and test data based on the reporting time of incident links. Specifically, we consider the data from January 01, 2022 to September 30, 2022 for training the models and the rest 3 months of data for evaluation. As majority of the linked incidents appear from the same service, we down sample related incident pairs coming from the same service, while prioritizing the related pairs coming from different services and workloads. For each of the positive pair of related incidents, we generate a negative sample (not related to anchor incident) that has reported within 4 hours of lookback time-window. For generating the dependency graph, we considered all the incident links from January 01, 2021 to September 30, 2022.

\subsubsection{Parameter configuration}
For generating the sub-graph of a service from the global dependency graph, we chose nodes that lie within 3 hop distance. We use node2vec \cite{grover2016node2vec} graph transformer with embedding dimension of 32, walk length l = 20, and number of walks r=100. For default experimental settings, we use a batch size of 200 and trained the model for 20 epochs. For the training network, we used 2 LSTM layer, each having 16 hidden nodes and the drop probability is set to 0.3. We perform a grid search between $[5e^{-4}, 1e^{-4},5e^{-3},1e^{-3}, 5e^{-2}]$ to identify the learning rate. For textual embeddings, we perform grid search between $[25, 50]$ for the dimension of title, topology, monitor ID, failure type and owning team, and the output dimension is searched from $[50,100, 200]$. For the graphical embeddings, we used grid search between $[16, 32, 48, 64]$ for identifying the input, hidden, output and text projection dimensions.  

Our experiments are conducted on Ubuntu 20.04 with Azure compute cluster having 4 nodes each with 24-core Intel Xeon E5-2690 v3 CPU with 224 GB memory and a single NVIDIA Tesla K80 GPU accelerator. On an average, it takes about 27 hours to complete 20 epochs of training for our methods on the cluster.

\subsubsection{Methods} We experimented with 3 version of DiLink and compared the performance with 2 baseline methods.

\paragraph{\textbf{Baseline. }} For the baseline method, we only used embeddings of textual information (i.e., title, topology, monitor ID, failure type and owning team) of incidents to generate the similarity scores.
\paragraph{\textbf{Concatenation. }} For this method, we generate the embeddings of textual and dependency graph, and concatenate those embeddings together to get the final embedding.  
\paragraph{\textbf{LiDAR \cite{chen2020identifying}. }} For this method, we employ two Siamese networks, one computes the similarity score from textual embeddings and other computes the similarity score from embeddings of localized sub-graphs. Finally, we took a convex combination (e.g., equal weights on both scores) of these scores to identify the similarity between two incidents.   
\paragraph{\textbf{DiLink-GCN. }} For DiLink, we used Orthogonal Procrustes method to project the text embeddings into graph embedding space. For DiLink-GCN, we used Graph Convolution Network \cite{kipf2016semi} for generating the final sub-graph embedding. 
\paragraph{\textbf{DiLink-GAT. }} For DiLink-GAT, we used Graph Attention Network \cite{velivckovic2017graph} for generating the final sub-graph embedding. 
\paragraph{\textbf{DiLink-GSAGE. }} For DiLink-GSAGE, we used GraphSAGE \cite{hamilton2017inductive} embedding generation algorithm for the final sub-graph embedding. 

\subsection{Experimental results}
\begin{figure*}[ht]
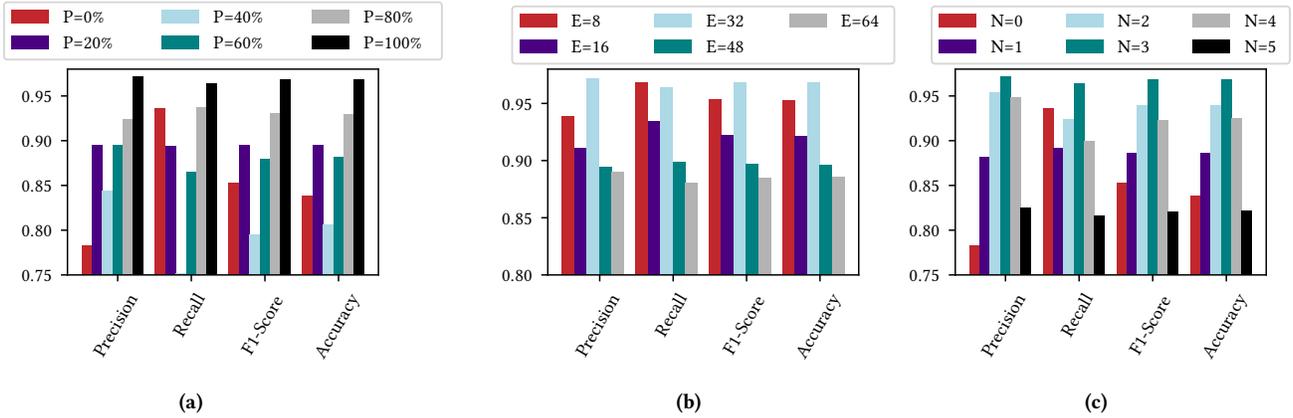

    \centering
    \begin{subfigure}{0.3\textwidth}
        \scalebox{0.7}{\input{Figures/subgraph_variation_v2.pgf}}
       \vspace{-4mm}  \caption{}
    \end{subfigure} \hskip 1.2cm
    \begin{subfigure}{0.3\textwidth}
        \scalebox{0.7}{\input{Figures/Embeddings_variation.pgf}}
          \vspace{-4mm} \caption{}
    \end{subfigure}
    \begin{subfigure}{0.3\textwidth}
        \scalebox{0.7}{\input{Figures/Neighbours_variation.pgf}}
          \vspace{-4mm} \caption{}
    \end{subfigure}
    \vspace{-0.15in}
\caption{Performance with varying (a) number of edges in dependency graph; (b) graph embedding sizes; and (c) number of neighbourhood hops.}
  \Description{Performance with varying (a) number of edges in dependency graph; (b) graph embedding sizes; and (c) number of neighbourhood hops.}
    \vspace{-0.1in}
  \label{fig:subgraph}
\end{figure*}
Since incident link prediction is a binary classification problem, we adopt widely-used classification
measures for evaluation: (1) \emph{Precision}: percentage of rightly predicted true links out of all predicted true related links; (2) \emph{Recall}: percentage of true links whose labels are rightly predicted; (3) \emph{F1-Score}: harmonic mean of precision and recall; and (4) \emph{Accuracy}: percentage of rightly predicted links. To evaluate the performance, we generated 5 test dataset each having around 40,000 incident pairs. The average and standard deviation in key metrics from inference results over these 5 test datasets for different approaches are shown in Table~\ref{tbl:performance}.

\paragraph{\textbf{Can dependency graph data improves incident linking performance?}} As expected, due to misalignment between embeddings of textual and graphical components, the performance of Concatenation method suffers. Even with additional dependency graph information, the F1-Score for the Concatenation method reduces by 2\%, while the Accuracy drops by more than 3\% over the baseline method that leverages only textual information. Separately learning the similarities from textual and graphical data with LiDAR approach improves the F1-score marginally over the baseline and concatenation method. 
With alignment of textual and graphical embeddings using Orthogonal Procrustes, we observe a significant gain in performance over both the baseline and concatenation method. DiLink-GCN improves the F1-Score and Accuracy by 12\% and 13\%, respectively over the baseline method. DiLink-GSAGE method further improve the accuracy of the predictions by around 1\% over the DiLink-GCN method. Overall, DiLink-GAT method provides the best performance, where the Precision, Recall, F1-Score and Accuracy is improved by roughly 20\%, 6\%, 14\% and 15\%, respectively over the baseline method.  

\paragraph{\textbf{How the linking accuracy varies between within-service, cross-service and cross-workload incident pairs?}} In Table~\ref{tbl:performance}, we provide the performance of different methods for incident pairs coming from same service, different service (but same workload) and from different workloads. We observe a consistent pattern across the board that the DiLink methods always outperforms the baseline and benchmark (i.e., concatenation and LiDAR) methods. For within service incidents, DiLink-GSAGE method achieves almost perfect Recall score, but having relatively lower precision score of 0.93. DiLink-GAT method provides the best performance among all the methods for incidents coming from the same service. For incident pairs coming from different services, but from same workload, DiLink-GSAGE method outperforms DiLink-GAT method in terms of Accuracy by 2\%. However, in terms of the F1-Scores, DiLink-GAT method still outperforms all the other approaches. For incident pairs coming from different workloads, where the distance between two services in dependency graph can be comparatively higher, DiLink-GSAGE method delineates the best performance with almost 1\% higher F1-Score than the DiLink-GAT method.

\paragraph{\textbf{How completeness of the dependency graph affect the linking accuracy?}} To investigate how the completeness of dependency graph impacts the performance, we randomly sample a fixed percentage of edges and generate an incomplete modified graph. In Figure~\ref{fig:subgraph}(a), we report the key performance metrics (i.e., Precision, Recall, F1-Score and Accuracy) by varying the number of edges in the dependency graph. As expected, when $P=0\%$, we provide an isolated node as sub-graph, which provides an F1-Score at par with baseline method that uses only textual data. On the other hand, when we consider all the edges in the dependency graph ($P=100\%$), we obtain the best performance with an F1-score of 0.96. As expected, sampling a subset of edges randomly from the original dependency graph degrades the performance as we ignore crucial dependency information among services.

\subsubsection{Parameter sensitivity}
For generating the sub-graphs for each service, we used two important tunable parameters: (1) embedding size of the input sub-graph; and (2) number of hops to consider for incoming and outgoing edges. We now provide a sensitivity analysis by varying these input parameters.

\paragraph{\textbf{Varying embedding size. }}In Figure~\ref{fig:subgraph}(b), we demonstrate the performance by varying the embedding size of input graph from 8 to 64, while fixing other parameter values that are obtained from grid search with hyperdrive experiments. With embedding size of 32, we obtain the best performance, which is used in default set of experiments.  

\paragraph{\textbf{Varying number of neighbourhood hops. }}In Figure~\ref{fig:subgraph}(c), we show the performance by varying number of neighbourhood hops ($N$) for generating the sub-graph from 0 to 5. When $N=0$, we obtain an isolated node for the sub-graph. On the other hand, when the value of $N$ increases, the size of sub-graph grows gradually which resembles with global dependency graph, and therefore, we loose critical localized information. We obtain the right trade-off in performance with $N=3$, which is employed for default experiments.

\section{Real-world Deployment} \label{deployment}

\begin{figure*}[t]
    \centering
    \includegraphics[width=0.7\textwidth]{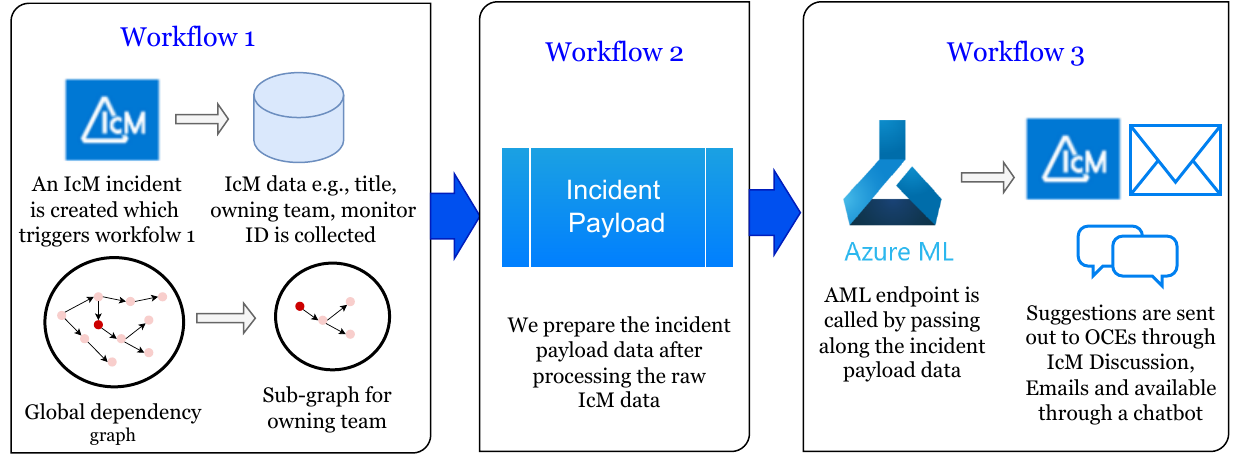}
    \vspace{-0.1in}
    \caption{High level end-to-end inference process.}
    \Description{High level end-to-end inference process.}
    \vspace{-0.15in}
    \label{fig:inference}
\end{figure*}

\begin{figure*}[ht]
    \centering
     \includegraphics[width=0.9\textwidth]{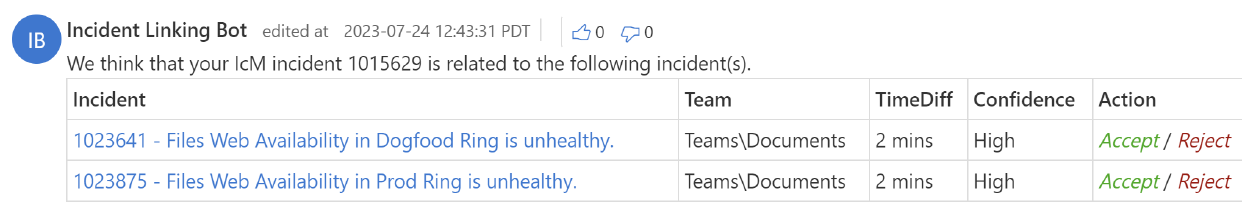}
    \vspace{-0.15in}
    \caption{Example link suggestions in IcM Discussion.}
    \Description{Example link suggestions in IcM Discussion.}
    \vspace{-0.1in}
    \label{fig:suggestions}
\end{figure*}

Given the significant boost in performance over baseline methods, we are currently in the final stage of deploying the DiLink model across 610 services from 5 workloads within \company{}. To the best of our knowledge, due to the sensitivity of the problem, no open-source dataset is publicly available with incident metadata and service dependency graph that can be used as benchmark for performance evaluation\footnote{To comply with confidentiality requirements, we also cannot release the dataset used for experimentation or publish the trained model publicly.}. 

For the production deployment, we first deploy our trained incident linking model using Azure machine learning (AML) platform and create an AML endpoint. For each newly created incident, we gather its textual information from the IcM portal through IcM automation, pass the features along with the sub-graph of incident owning team to the AML endpoint, and generate the final embeddings. We compare each incoming incident's embeddings to the embeddings of all incidents that were created in the last $\Delta t$ lookback time-window to calculate the embedding distances. We set the value of lookback period $\Delta t$ to the $90^{th}$ percentile value of create time difference between pairs of historically linked incidents, which is 4 hours. If the embedding distances are lower than the predefined threshold value found from model training, the pairs are considered as positive links.

We leverage AML to train our model, store the model artifacts, and to deploy the model for real-time endpoint creation. The AML real-time endpoint is deployed on an Azure Kubernetes (AKS) cluster. The AKS cluster has benefits such as auto-scaling, auto-upgrading for OS versions, integrated logging and monitoring, and ability to handle enterprise-grade workloads. 
The AML endpoint performs the following functions: 
(a) Generate embeddings for an incoming incident with the relevant dependency sub-graph; (b) Store the embeddings data in a database; (c) Runs a Kusto query to calculate distances between pairs of incidents; and (d) Return as response the pairs of incidents that should be linked and sent to OCEs. 
During inference, we leverage the real-time AML endpoint and the automation functionalities within our IcM system that are built using Azure Logic Apps. IcM Automation natively has connectors to IcM data, which assists us to easily collect incident data upon incident creation. At a high level, the following 3 IcM workflows are created for inference (see Figure~\ref{fig:inference} for more details):
(a) Workflow 1: Triggered when an incident is created and collects the incident data from IcM;
(b) Workflow 2: Triggered by Workflow 1, and prepares the incident payload to pass to the AML endpoint; and
(c) Workflow 3: Triggered by Workflow 2. It calls the AML endpoint with the incident payload, predicts related links and sends the link suggestions to OCEs.

We communicate the link suggestions with the OCEs to assist in finding root causes, finding related issues, or joining the incident bridge faster through three channels: IcM's discussion section, emails, and Teams chatbot. An example of link suggestions sent to OCEs through IcM Discussion is shown in Figure~\ref{fig:suggestions}. OCEs are presented the link suggestions along with relevant data on the incidents (e.g., incident title, owning team, create time difference, model confidence). For each incident suggestion, we also add feedback buttons that OCEs can use to accept or reject the link upon review.
If an OCE accepts a link suggestion, the link is stored in IcM, and we count that link as a True Positive (TP). If an OCE rejects a link suggestion (i.e., False Positive (FP)), they are presented with another page asking for short justification so as to leverage their expertise and domain knowledge for continuous model improvement. 

\section{Lessons Learned and Threats} \label{lessons}

\paragraph{\textbf{Fusing information from different sources. }}
As demonstrated in experimental results, both the textual description and the dependency structure contain critical information about different aspects of incident links, which motivates the need to combine these two modality of information together in our model. The LSTM layers can capture the underlying semantic information from textual data such as incident title, topology and detection details, and the Siamese network is capable of learning to match these semantic details. Moreover, with efficient graph transformer and embeddings (e.g., graph attention network), we can encapsulate the relationship between different teams and services, which helps to declutter the complex relationship between cross-team incidents. By integrating both textual and dependency information together, our proposed method enjoys a great improvement in the incident linking task.

\paragraph{\textbf{Modeling information from dependency graph. }}
Construction and processing of dependency graph plays a vital role in improving the performance of incident linking model. We observe that while service metadata provides crucial dependency information, it might not provide a complete dependency graph (as identifying dependencies among teams is a non-trivial problem in hyperscale services) and therefore, adding dependency links from historical incident relationships greatly improve the performance. In addition, generating sub-graphs with right number of neighbour nodes is vital to obtain the right trade-off between providing minimal information (without any neighbours) and exposing to generalized graph (with high number neighbourhood hops).

\paragraph{\textbf{Alignment of multi-modal data embeddings. }}
Ensembling information from different data modality is a challenging problem. For incident linking, a simple concatenation of textual and graphical embeddings leads to poor performance due to misalignment of embedding dimensions. We exhibit that projecting text embeddings to graphical embedding space using Orthogonal Procrustes and then concatenating the two sets of embeddings significantly boost the performance of incident linking task. 

\paragraph{\textbf{Threat to validity. }} 
There are several threats to our study. As the labels used as ground truth are mostly provided by OCEs, there is a potential risk of positive related links are noisy. However, as the OCEs have expertise and domain knowledge to understand the relationship between incidents and the links are usually validated by multiple OCEs, the potential risk is minimal but inevitable. On the other hand, there could be missing links in our database. Due to limited time and high priority incidents, OCEs might miss to add all the positive links, so our negative samples could be noisy as well. To minimize this risk, we randomly picked negative samples from a pool of incidents reported within 4 hours of window. 
We train and evaluate the models with incident data from 5 workloads within our organization.  Our insights and experimental results may not be generalized across all the services within \company{} that we have not considered in our study and may not represent the behaviour of other public cloud services. 
However, as these 5 workloads run hyperscale services continuously used by hundred of millions of users, our incident dataset represents the typical nature and challenges of large-scale cloud services. Lastly, we have only considered machine reported incidents in our study, so the performance may not be replicated for identifying links between customer reported incidents. 
\section{Related Work} \label{related}
\subsection{Incident Management} Incident management is key to effectively and efficiently operate large-scale cloud services. There are several challenges in the various phases of the incident life-cycle \cite{shetty2021neural, azad2022picking, gao2020scouts, ghosh2022fight, bansal2019decaf, ahmed2023recommending, vaibhav2023detection} which have been a prominent subject of study within the Systems and Software Engineering communities. Chen \etal \cite{EmpiricalIcMICSE2019} found that mis-triaging can lead up to 10X delay in mitigation. To solve this challenge, Pham \etal \cite{pham2020deeptriage} proposed DeepTriage which leverages an ensemble model to route incidents to the right owning team. Despite the promising results, the authors call out several challenges in automated incident triaging, such as cold start problem, noisy data, lack of sufficient signals. Incident diagnosis has also been extensively studied to help minimize incident impact and the manual effort by the OCEs. Given the recent emergence of Large-language models like GPT, they have also been used for incident root-causing with promising results \cite{ahmed2023recommending, chen2023empowering}. In this work, we explore the problem of incident linking which would help accelerate all the phases of the incident life-cycle. For instance, the OCEs investigating the related incidents, can assist in identifying the right team for a newly reported incident.  Similarly, identifying linked incidents would accelerate root-causing and mitigation by avoiding siloed investigations.  

\subsection{Duplicate bug detection} 
Our work is most closely related to the significant amount of prior work on identifying duplicate bug reports \cite{gu2021bugs, sun2010discriminative, tian2012improved, hindle2019preventing}. Sun \etal \cite{sun2010discriminative} investigated the usage of a discriminative model to help identify duplicate bug reports. Hindle \etal \cite{hindle2019preventing} proposed a simplified way of avoiding duplicate bug reports by continuously retrieving related bug reports as the user is creating a bug report. Zhou \etal \cite{zhou2012learning} leveraged learning to rank approach along with textual and statistical features of bug reports to identify duplicate bug reports. In this work, we tackle a more complex problem of identifying linked incidents in a distributed micro-service architecture setting. Here, we need to identify incident links across thousands of services. Also, our notion of incident \textit{links} is more abstract and context dependent than the prior work on \textit{duplicate} bug report identification. 

There has been limited work in the incident management which has looked into the problem of incident linking \cite{chen2020identifying,chen2022online,gu2020efficient}. Gu \etal \cite{gu2020efficient} propose LinkCM which leverages transfer learning to link customer reported incidents to monitor reported ones. Chen \etal \cite{chen2020identifying} proposed the LiDAR framework which leverages the textual description and the component dependency graph for predicting the incident links. They obtained the similarity scores from two Siamese networks (one for text and another for graph data) and employed a convex combination of these two scores. In contrast, we look into the problem of linking MRIs by leveraging service dependency graph jointly with textual information (after aligning both the embeddings together) so as to accurately link the incidents across different services and workloads for minimizing the manual effort and accelerating the incident resolution process.

\section{Conclusion} \label{conclusion}
Identifying similar incident links is an important and challenging task for building an intelligent IcM system and for ensuring reliable large-scale cloud services. In this paper, we investigate the linked incidents from 610 services within \company{} and exhibit that leveraging dependency information among services is crucial for incident linking task. We propose to augment the textual information with dependency graph data to capture important aspects of incident links. To combine the textual and graphical data efficiently, we leverage Orthogonal Procrustes method for projecting text embeddings to graph embedding space. Extensive experimental results demonstrate that our proposed DiLink model provides significant gain in performance over state-of-the-art benchmark methods.
In future, this work can be extended in two ways: (1) evaluate the efficacy and challenges for employing DiLink model for predicting links between customer and machine reported incidents; and (2) extending current model for solving other incident management tasks such as triaging and root causing of incidents.

\bibliographystyle{ACM-Reference-Format}
\bibliography{references}

\end{document}